\documentclass[aps,prl,twocolumn,showpacs,amsmath,amssymb]{revtex4-2}
\usepackage{amsmath}
\usepackage{graphicx}
\usepackage{subfigure}
\usepackage{epstopdf}
\usepackage{color}
\usepackage{makecell}

\usepackage{multirow}
\usepackage{setspace}
\usepackage{overpic}
\usepackage{amssymb}
\usepackage[bookmarksnumbered, pdfstartview=FitH,colorlinks,urlcolor=blue, citecolor=blue,linkcolor=blue] {hyperref}
\usepackage{lineno}
\usepackage{bm}
\usepackage{rotating}
\usepackage[utf8]{inputenc}

\let\oldequation\equation
\let\oldendequation\endequation

\renewenvironment{equation}
  {\linenomathNonumbers\oldequation}
  {\oldendequation\endlinenomath}

\begin{document}

\title{{\bf \boldmath Observation of $D^+\to\eta^\prime\mu^+\nu_\mu$ and First Experimental Study of $D^+\to \eta^\prime \ell^+\nu_\ell$ Decay Dynamics}}

\author{
M.~Ablikim$^{1}$, M.~N.~Achasov$^{4,c}$, P.~Adlarson$^{76}$, O.~Afedulidis$^{3}$, X.~C.~Ai$^{81}$, R.~Aliberti$^{35}$, A.~Amoroso$^{75A,75C}$, Q.~An$^{72,58,a}$, Y.~Bai$^{57}$, O.~Bakina$^{36}$
% ...
\\
\vspace{0.2cm}
(BESIII Collaboration)\\
\vspace{0.2cm} {\it
$^{1}$ Institute of High Energy Physics, Beijing 100049, People's Republic of China\\
% ...
$^{3}$ Bochum  Ruhr-University, D-44780 Bochum, Germany\\
$^{4}$ Budker Institute of Nuclear Physics SB RAS (BINP), Novosibirsk 630090, Russia\\
% ...
$^{35}$ Johannes Gutenberg University of Mainz, Johann-Joachim-Becher-Weg 45, D-55099 Mainz, Germany\\
$^{36}$ Joint Institute for Nuclear Research, 141980 Dubna, Moscow region, Russia\\
% ...
$^{57}$ Southeast University, Nanjing 211100, People's Republic of China\\
$^{58}$ State Key Laboratory of Particle Detection and Electronics, Beijing 100049, Hefei 230026, People's Republic of China\\
% ...
$^{72}$ University of Science and Technology of China, Hefei 230026, People's Republic of China\\
% ...
$^{75}$ University of Turin and INFN, (A)University of Turin, I-10125, Turin, Italy; (B)University of Eastern Piedmont, I-15121, Alessandria, Italy; (C)INFN, I-10125, Turin, Italy\\
$^{76}$ Uppsala University, Box 516, SE-75120 Uppsala, Sweden\\
% ...
$^{81}$ Zhengzhou University, Zhengzhou 450001, People's Republic of China\\
\vspace{0.2cm}
$^{a}$ Deceased\\
% ...
$^{c}$ Also at the Novosibirsk State University, Novosibirsk, 630090, Russia\\
% ...
}
}

\begin{abstract}

Using $20.3\,\rm fb^{-1}$ of $e^+e^-$ collision data collected
at the center-of-mass energy 3.773\,GeV with the BESIII detector, we report the first observation of the semileptonic decay  $D^+\to \eta^\prime \mu^+\nu_\mu$ with significance of $8.6\sigma$ including systematic uncertainties, and 
an improved measurement of $D^+\to \eta^\prime e^+\nu_e$. 
The branching fractions of $D^+\to \eta^\prime \mu^+\nu_\mu$ and $D^+\to \eta^\prime e^+\nu_e$ are determined to be
$(1.92\pm0.28_{\rm stat}\pm 0.08_{\rm syst})\times 10^{-4}$ and
$(1.79\pm0.19_{\rm stat}\pm 0.07_{\rm syst})\times 10^{-4}$, respectively. 
The ratio of the two branching fractions is determined to be ${\mathcal R}_{\mu/e}=1.07\pm0.19_{\rm stat}\pm 0.03_{\rm syst}$, which agrees with the theoretical expectation of lepton flavor universality within the Standard Model.
From an analysis of the $D^+\to \eta^\prime \ell^+\nu_\ell$ decay dynamics, the product of the hadronic form factor $f_+^{\eta^{\prime}}(0)$ and the CKM matrix element $|V_{cd}|$ is measured for the first time, giving $f^{\eta^\prime}_+(0)|V_{cd}| = (5.92\pm0.56_{\rm stat}\pm0.13_{\rm syst})\times 10^{-2}$. 
The $\eta-\eta^\prime$ mixing angle in the quark flavor basis is determined to be
$\phi_{\rm P} =(39.8\pm0.8_{\rm stat}\pm0.3_{\rm syst})^\circ$.
\end{abstract}

\maketitle

\oddsidemargin  -0.2cm
\evensidemargin -0.2cm

In the standard model (SM), semileptonic (SL) $D$ decays involve the interaction of a leptonic current with a hadronic current~\cite{Ivanov:2019nqd}. 
By analyzing the differential decay rates of SL $D$ decays, one can determine the hadronic form factors (FFs) 
and the Cabibbo-Kobayashi-Maskawa (CKM) matrix~\cite{Cabibbo:1963yz,Kobayashi:1973fv} elements $|V_{cd(s)}|$, which parameterize the strong and weak interaction effects, respectively.
The FFs may be calculated by lattice quantum chromodynamics (QCD) or other, less-rigorous methods and models. Measurements of the FFs of the SL $D$ decays are important to test these methods in the charm sector.

In general, SL $D^{0(+)}$ decays into various ground state mesons have been well-explored~\cite{PDG2024,Ke:2023qzc}. However, no experimental study of $D^+\to\eta^\prime\mu^+\nu_\mu$ exists.  A study of the $D^+\to\eta^\prime\ell^+\nu_\ell$  ($\ell=\mu$ or $e$) 
decay dynamics  provides an opportunity to determine the $D^+\to \eta^\prime$ hadronic FF $f_+^{\eta^{\prime}}(q^2)$~\cite{Duplancic:2015zna,Offen:2013nma,Soni:2018adu,Ivanov:2019nqd,Hu:2023pdl,Faustov:2019mqr} and the CKM matrix element $|V_{cd}|$. Here, $q^2$ is the square of the four-momentum transferred to the $\ell^+\nu_\ell$ system.
Predictions for $f^{\eta^{\prime}}_+(0)$ vary from 0.292 to 0.76, based on theoretical models such as the
 QCD light-cone sum rules (LCSR)~\cite{Offen:2013nma,Duplancic:2015zna}, the covariant confined quark model (CCQM)~\cite{Soni:2018adu,Ivanov:2019nqd}, the relativistic quark model (RQM)~\cite{Faustov:2019mqr}, and the light cone harmonic oscillator model (LCHO)~\cite{Hu:2023pdl}. 
Verification of charm FF calculations would also help to constrain the calculations of the FFs of SL $B$ decays, which provide precise determinations of the CKM matrix elements $|V_{cb}|$ and $|V_{ub}|$~\cite{Koponen:2012di,Koponen:2013tua,Brambilla:2014jmp,Bailey:2012rr}.

In the SM, lepton flavor universality (LFU) requires the same couplings between the three families of leptons and the gauge bosons. 
In recent years, there have been reported hints of LFU violation in $B\to \bar D^{(*)}\ell^+\nu_\ell$~\cite{HFLAV,babar_1,babar_2,lhcb_1,belle2015,belle2016,Belle:2019rba,LHCb:2023zxo} and the anomalous magnetic moment of the muon~\cite{Muong-2:2006rrc,Muong-2:2021ojo}.
A measurement of the
ratio of the branching fractions (BFs) of $D^+\to\eta^{\prime}\mu^+\nu_\mu$ and $D^+\to\eta^{\prime} e^+\nu_e$ (${\mathcal R}_{\mu/e}$) would thus offer an important complementary test of LFU in
the charm sector, with the current SM prediction being in the range of
$0.94-0.95$~\cite{Soni:2018adu,Cheng:2017pcq,Faustov:2019mqr}.
In addition, the singlet-octet $\eta$-$\eta^{\prime}$-gluon mixing angle $\phi_{\rm P}$~\cite{Christ:2010dd,Dudek:2011tt}, related to the QCD anomaly and the breaking of chiral symmetry~\cite{Chao:1990im,Gershtein:1976mv}, can be determined via $\cot^4\phi_{\rm P}=\frac{\Gamma_{D^+_s\to\eta^\prime \ell^+\nu_\ell}/\Gamma_{D^+_s\to\eta
    \ell^+\nu_\ell}}{\Gamma_{D^+\to\eta^\prime \ell^+\nu_\ell}/\Gamma_{D^+\to\eta \ell^+\nu_\ell}}$~\cite{DiDonato:2011kr}. Here, differences in the phase space and the FFs for $D^+$ and $D_s^+$, as
well as the gluonium component of the $\eta^\prime$ meson, cancel in the ratio,
allowing for an accurate determination of $\phi_P$, which in turn can
help discriminate between different treatments of the relevant
non-perturbative QCD effects~\cite{Cao:2012nj,Ke:2010htz}.

Previously, CLEO and BESIII reported the BF of $D^+\to\eta^{\prime}e^+\nu_e$~\cite{CLEO:2010pjh,BESIII:2018eom} with large uncertainties, of order $25\%$.  This Letter reports the first observation of $D^+\to \eta^\prime \mu^+\nu_\mu$, the improved measurement of $D^+\to \eta^\prime e^+\nu_e$, and the first analysis of $D^+\to \eta^\prime \ell^+\nu_\ell$ decay dynamics by analyzing 20.3~fb$^{-1}$~\cite{BESIII:2024lbn} of $e^+e^-$ collision data taken at the center-of-mass energy ($E_{\rm CM}$) of 3.773 GeV with the BESIII
detector.  The dataset is about seven-times larger compared with Ref.~\cite{BESIII:2018eom}, and we now also investigate potential $\mu$-$e$ LFU in the $D^+\to \eta^{\prime}\ell^+\nu_\ell$ decays
in the full kinematic range across four $q^2$ intervals~\cite{charge}. 

Details about the design and performance of the BESIII detector are given in Refs.~\cite{BESIII:2009fln,BESIII:2020nme}.
Simulated samples produced with a {\sc
Geant4}-based~\cite{GEANT4:2002zbu} Monte Carlo (MC) package, which
includes the geometric description of the BESIII detector and the
detector response, are used to determine the detection efficiency
and to estimate backgrounds. The simulation includes the beam
energy spread and initial state radiation (ISR) in the $e^+e^-$
annihilations modeled with the generator {\sc
kkmc}~\cite{kkmc}.
Simulation samples of the $D^+\to\eta^\prime\ell^+\nu_\ell$ signal process are
generated using the two-parameter series-expansion model as described
in Ref.~\cite{Chikilev:1999zn}, with the parameter values estimated \textit{in situ} in this
analysis.
The background is studied using an inclusive simulation sample that consists of the
production of $D\bar{D}$
pairs with consideration of quantum coherence for all neutral $D$
modes, the non-$D\bar{D}$ decays of the $\psi(3770)$, the ISR
production of the $J/\psi$ and $\psi(3686)$ states, and the
continuum processes incorporated in {\sc kkmc}.
The known decay modes are modeled with {\sc evtgen}~\cite{evtgen} using the known BFs taken from the
Particle Data Group~\cite{PDG2024}, while the remaining unknown decays
from the charmonium states are modeled with {\sc
lundcharm}~\cite{lundcharm}. Final state radiation
from charged final state particles is incorporated with the {\sc
photos} package~\cite{photos}.

The $D^+\to \eta^\prime \ell^+\nu_\ell$ candidates are selected in events with $D^-$ decays in one of the six decay modes $D^-\to K^+\pi^-\pi^-$, $K^0_S\pi^-$, $K^+\pi^-\pi^-\pi^0$, $K^0_S\pi^-\pi^0$, $K^0_S\pi^+\pi^-\pi^-$, and $K^0_S K^-$. A reconstructed $D^-$ meson candidate is referred to as a single-tag (ST) candidate. An event in which a signal $D^+\to \eta^\prime \ell^+\nu_\ell$ decay candidate and an ST $D^-$ are simultaneously found is referred as a double-tag (DT) event.
 The BF of $D^+\to \eta^\prime \ell^+\nu_\ell$ is determined  by
\begin{equation}
\label{eq:br}
{\mathcal B} = \frac{N_{\rm DT}}{N_{\rm ST}^{\rm tot}\epsilon_{\rm sig}},
\end{equation}
where $N_{\rm ST}^{\rm tot}=\sum_{i=1}^6{N_{\rm ST}^i}$ and
$N_{\rm DT}$ are the total ST and DT yields summing over tag mode $i$, 
$\epsilon_{\rm sig}=\sum_{i=1}^{6}{\frac{N_{\rm ST}^{i}}{N_{\rm ST}^{\rm tot}}\frac{\epsilon_{\rm DT}^{i}}{\epsilon_{\rm ST}^{i}}}$ is the effective signal efficiency of selecting $\eta^{\prime}\ell^+\nu_\ell$ in the presence of ST $D^-$, where $\epsilon_{\rm ST}^{i}$ and  $\epsilon_{\rm DT}^{i}$ are the ST and DT efficiencies for the $i^{\rm th}$ tag mode, respectively.

The selection criteria of $\pi^\pm$, $K^\pm$, $K_S^0$, photon, and $\pi^0$ candidates of both tag and signal sides are the same as Refs.~\cite{BESIII:2023exq,BESIII:2024slx}.
The tagged $D^-$ mesons are selected using two variables, the energy difference
$\Delta E \equiv  E_{D^-} - E_{\rm beam}$
and the beam-constrained mass
$M_{\rm BC} \equiv \sqrt{E^{2}_{\rm beam}-|\vec{p}_{D^-}|^{2}}$,
where $E_{\rm beam}$ is the beam energy, and $\vec{p}_{D^-}$ and $E_{D^-}$ are the momentum and the energy of the $D^-$ candidate in the $e^+e^-$ rest frame.
If there are multiple combinations in an
event, the combination with the minimum $|\Delta E|$ is chosen for each tag
mode. 
The ST yields and efficiencies are determined by fitting the $M_{\rm BC}$  distributions of the accepted candidates in data and in the inclusive simulation sample, respectively. 
Details about the fits can be found in Ref.~\cite{BESIII:2024slx}, and the obtained fit results, ST yields and efficiencies are shown in Ref.~\cite{Supplement}.
Summing over all six tag modes, the total ST yield is $N^{\rm tot}_{\rm ST}=(10289.8\pm3.7)\times10^3$.

In the presence of the ST $D^-$, the $\eta^\prime\to\eta\pi^+\pi^-$, $\eta^\prime\to\gamma\pi^+\pi^-$ and $\ell^+$ candidates of signal side are selected from the remaining tracks and showers, with the same selection criteria as Refs.~\cite{BESIII:2023ajr,BESIII:2023gbn} except below.
 The $\eta\to\gamma\gamma$ candidates are required to have $\gamma\gamma$ invariant mass between $(0.505,\,0.575)$~GeV$/c^{2}$. The radiative photon coming from $\eta^\prime\to\gamma \pi^+\pi^-$ cannot form a $\pi^0$ with any unused photons.
 To reduce backgrounds due to particle misidentifications, the positron candidate is further required to have a deposited energy in the electromagnetic
calorimeter (EMC) greater than 0.8 times its momentum measured by the drift chamber. 
For muon candidate, the deposited energy in the EMC is further required to be less than 0.3\,GeV; and its particle identification (PID) likelihood, combined from the time-of-flight information, the specific ionization energy loss measured in the drift chamber, and the shower information of the EMC, for the muon hypothesis is additionally required to be greater than that for pion.

The missing energy and momentum of the neutrino of the signal SL decay are derived
as $E_\nu \equiv E_{\rm CM} - \Sigma_i E_i$ and $\vec{p}_\nu \equiv -\Sigma_i \vec{p}_{i}$, respectively, where
$E_i$ and $\vec{p}_i$ are the energy and momentum of the particle $i$, with $i$ running over
the ST $D^-$, the $\eta^{\prime}$ and $\ell^+$ of the signal side.
The yield of the SL signal events is determined by a fit to the distribution of the kinematic variable
$U_{\rm miss} \equiv E_\nu/c^2 - |\vec{p}_\nu|/c$.
To improve the resolution, the candidate tracks, along with the neutrino, are subjected to a three-constraint kinematic fit requiring energy and momentum conservation, constraining the invariant mass of $\eta\pi^+\pi^-$ or $\gamma\pi^+\pi^-$ to the known $\eta^\prime$ mass~\cite{PDG2024}, and constraining the invariant
mass of the daughter particles of each $D^\pm$ to the known $D^\pm$ mass~\cite{PDG2024}. In the case of multiple combinations, the one giving the minimum $\chi^2$ is kept for the further analysis. The averaged multiplicities are 1.21 and 1.08 for $D^+\to\eta^\prime \mu^+\nu_\mu$ and  $D^+\to\eta^\prime e^+\nu_e$, respectively.
The $\chi^2$ is required to be less than 40 (200), which is obtained by the optimization with the simulation sample, for $D^+\to\eta^\prime \mu^+\nu_\mu$ ($\eta^\prime e^+\nu_e$) to further suppress the non-$D^+D^-$ backgrounds.  

The DT candidates are vetoed if they contain any additional charged tracks ($N_{\rm extra}^{\rm char}$) or $\pi^0$ candidates reconstructed from two unused photons ($N_{\rm extra}^{\pi^0}$). The selection criteria of additional charged tracks and $\pi^0$ candidates are the same as Refs.~\cite{BESIII:2023exq,BESIII:2024slx}. Furthermore, the energy of any unused shower ($E_{\rm \gamma~extra }^{\rm max}$) in an event is required to be less  than 0.3~GeV to suppress the backgrounds with extra neutral particles.
To further reject the peaking backgrounds from $D^+\to\eta^{\prime} \pi^+$ caused by the $\ell$-$\pi$ misidentification,
the invariant masses of $\eta^{\prime} \mu^+(\eta^{\prime}e^+)$, $M_{\eta^{\prime} \mu^+(\eta^{\prime}e^+)}$, are required to be less than $1.72~(1.80)$~GeV/$c^2$.
The opening angle between the missing momentum and the most energetic unused shower $\theta_b$ is required to satisfy $\cos\theta_b<0.70~(0.86)$ for $D^+\to\eta^\prime \mu^+\nu_\mu$ ($\eta^\prime e^+\nu_e$).
For $D^+\to\eta^\prime_{\gamma\pi^+\pi^-}\ell^+\nu_\ell$, the $\eta^\prime\to\gamma\pi^+\pi^-$ is actually dominated by $\eta^\prime\to\gamma\rho^0$~\cite{BESIII:2017kyd}, the decay helicity angle of the daughter pion in the rest frame of $\rho^0$, $\theta_{\pi,\rho}$, is required  to satisfy $|\cos\theta_{\pi,\rho}|<0.82(0.86)$ for $D^+\to\eta^\prime \mu^+\nu_\mu$ ($\eta^\prime e^+\nu_e$). 
The above requirements of $E_{\rm \gamma~extra }^{\rm max}$, $M_{\eta^{\prime} \mu^+(\eta^{\prime}e^+)}$, $\cos\theta_b$, and $\cos\theta_{\pi,\rho}$ were optimized using the simulated samples.

To suppress the backgrounds due to the final state radiation, the angle between the direction of the decay photon and the positron momentum is required to be greater than 0.20 radians.
To reject the backgrounds arising from  $D^+\to\pi^+\pi^-\ell^+\nu_\ell$ with $\pi^+\pi^-$ from $\rho^0$ or $K_S^0$, $D^+\to\pi^+\pi^+\pi^-\pi^0$, and $D^+\to K_L^0\pi^+\pi^+\pi^-$, we use the recoil mass-squared, 
\begin{equation}
M_{\rm rec}^2=(E_{\rm CM}-E_{D^-}-E_{\pi^+\pi^-\ell^+})^2-(-\vec{p}_{D^-}-\vec{p}_{\pi^+\pi^-\ell^+})^2.
\end{equation}
In the $M_{\rm rec}^2$ distribution, the $D^+\to\pi^+\pi^-\ell^+\nu_\ell$, $D^+\to\pi^+\pi^+\pi^-\pi^0$, and $D^+\to K_L^0\pi^+\pi^+\pi^-$ candidates concentrate around 0.00, 0.02, and 0.25 GeV$^2$/$c^4$, respectively. To suppress these backgrounds, the events with 
 $M_{\rm rec}^2<0.075$~GeV$^2$/$c^4$ or  $M_{\rm rec}^2\in(0.211,\,0.308)$~GeV$^2$/$c^4$ for $D^+\to\eta^\prime \mu^+\nu_\mu$ and $M_{\rm rec}^2<0.050$~GeV$^2$/$c^4$ for  $D^+\to\eta^\prime e^+\nu_e$ are rejected. These requirements correspond to about $\pm 5\sigma$ around each peak.

\begin{figure}[htp]
\centering
\includegraphics[width=0.475\textwidth]{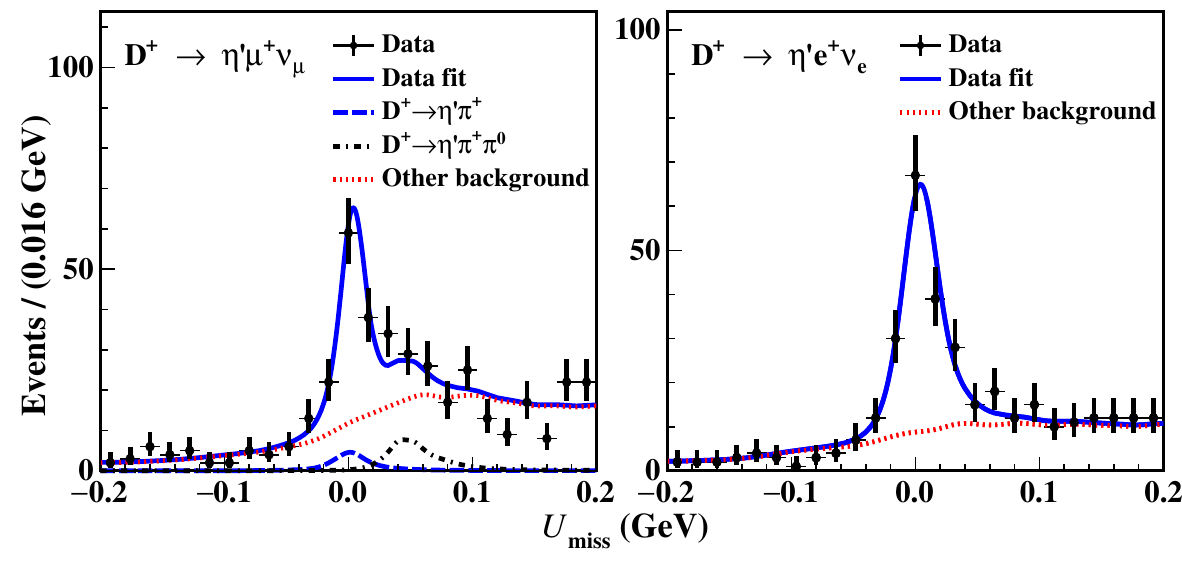}
\caption{Fits to the $U_{\rm miss}$ distributions of the accepted candidates for $D^+\to \eta^{\prime}\ell^+\nu_\ell$.
The points with error bars are data, the blue solid lines are the best fits,
the blue dashed (black dashed-dotted) lines are the peaking backgrounds $D^+\to\eta^{\prime} \pi^+(\eta^\prime \pi^+ \pi^0)$ , and 
the red dotted lines are the other backgrounds.
\label{fig:signal_yields_fromdata}}
\end{figure}

After imposing all above selection criteria, the resulting $U_{\rm miss}$ distributions of $D^+\to\eta^\prime \ell^+\nu_\ell$ of the accepted candidates summing over two $\eta^\prime$ reconstruction modes are exhibited in Fig.~\ref{fig:signal_yields_fromdata}.
The signal yields are extracted from an unbinned maximum likelihood fit
to these spectra.
The signal shape is taken from the signal simulation sample convolved with a Gaussian function with free parameters to
account for the resolution. The shapes of the peaking backgrounds of $D^+\to\eta^{\prime}\pi^+(\eta^{\prime}\pi^+\pi^0)$ and other backgrounds dominated by the $D\bar D$ decays~($\sim 90\%$) are modeled by the individual simulated shapes taken from the inclusive simulation sample.  The sizes of the peaking backgrounds are fixed to the expected yields from simulation, and the yields of other backgrounds are left free. 
The obtained signal efficiencies, signal yields, significances, and BFs are shown in Table~\ref{table:br}. 
The signal significances for $D^+\to\eta^\prime\ell^+\nu_\ell$ are calculated from the changes in likelihood between the nominal fits, shown in Fig.~\ref{fig:signal_yields_fromdata}, and the fits with the yields fixed to zero. Here, the likelihood distributions are estimated by the Bayesian approach after incorporating the systematic uncertainties~\cite{Stenson:2006gwf,cov:sys}.

   \begin{table}[hbtp]\small
   \centering
    \caption{Signal efficiencies $\epsilon_{\rm sig}$, signal yields ($N_{\rm DT}$), significances, and BFs ($\mathcal{B}$).  Efficiencies  include the BFs of the $\eta^{(\prime)}$ sub-decays.
The first and second uncertainties are statistical and systematic, respectively.  \label{table:br}}
 \begin{tabular}[t]{c|cc|cc}\hline\hline

Decay&\multicolumn{2}{c|}{$\eta^\prime \mu^+\nu_\mu$}&\multicolumn{2}{c}{$\eta^\prime e^+\nu_e$}\\
\hline
$\eta^{\prime}$ decay&$\eta\pi^+\pi^-$&$\gamma\pi^+\pi^-$&$\eta\pi^+\pi^-$&$\gamma\pi^+\pi^-$\\
$\epsilon_{\rm sig}$~(\%)&1.77$\pm$0.01&2.77$\pm$0.01& 2.70$\pm$0.01&5.50$\pm$0.01\\
$N_{\rm DT}$& \multicolumn{2}{c|}{90$\pm$13}& \multicolumn{2}{c}{151$\pm$16} \\
Significance&\multicolumn{2}{c|}{$8.6\sigma$}&\multicolumn{2}{c}{$12.9\sigma$}\\
$\mathcal{B}$~($\times10^{-4}$)&\multicolumn{2}{c|}{${1.92\pm0.28\pm 0.08}$}&\multicolumn{2}{c}{${1.79\pm0.19\pm0.07}$}\\

\hline\hline
        \end{tabular}
\end{table}

The systematic uncertainties in the BF measurements are listed in Table 2 of Ref.~\cite{Supplement} and discussed below.
The uncertainty in the ST $D^{-}$ yield is due to the fit to the
 $M_{\rm BC}$ distributions and is studied 
by varying the signal and background shapes. 
The uncertainties in the tracking or PID efficiencies of $\pi^\pm$, $e^+$, and $\mu^+$ are studied with the
control samples of DT hadronic events of $D^0\to K^-\pi^+, K^-\pi^+\pi^0, K^-\pi^+\pi^+\pi^-$, $D^+\to K^-\pi^+\pi^+$, $e^+e^-\to\gamma e^+e^-$, and $e^+e^-\to \gamma \mu^+\mu^+$, respectively. 
The uncertainties of the photon and $\pi^0$ reconstruction are assigned 
by studying the control samples of  $J/\psi\to \pi^0\pi^+\pi^-$~\cite{Ablikim:2011kv} and $D^0\to K^-\pi^+\pi^0$, respectively. Due to limited statistics, the uncertainty for $\eta$ reconstruction is assigned as equal to that for $\pi^0$ reconstruction since both are reconstructed from $\gamma\gamma$ decays. 
The uncertainties from the $\eta^{\prime}$ mass windows are estimated 
by analyzing the control sample of $J/\psi\to \phi\eta^{\prime}$.

The efficiencies of the  $M_{\eta^{\prime}\ell^+}$ and $\cos\theta_{\pi,\rho}$  requirements are greater than 95\% and the differences of these efficiencies between data and simulation sample are negligible.
The uncertainties of the $E_{\rm \gamma~extra }^{\rm max}$, $N_{\rm extra}^{\pi^0}$, and
 $N_{\rm extra}^{\rm char}$ requirements are analyzed with DT events of $D^+\to\eta^{\prime}\pi^+$.
 The uncertainties of the $\chi^2$ and $\cos\theta_b$ requirements are studied with the DT events of $D^+\to K_S^0\pi^0 e^+\nu_e$.
The uncertainties of the $M_{\rm rec}^2$ requirements are studied with 
the DT events of $D^+\to K_S^0\pi^0 e^+\nu_e$ and $D_s^+\to\eta^\prime_{\gamma\pi^+\pi^-}e^+\nu_e$.

The uncertainties due to the signal simulation model are estimated 
by comparing the DT efficiencies obtained by modified pole model~\cite{Becirevic:1999kt}. The uncertainties in the $U_{\rm miss}$ fit are studied  
by varying the sizes of the fixed peaking background yields and the top ten other backgrounds by varying the corresponding BFs by $\pm1 \sigma$~\cite{PDG2024}, varying the fraction of the $e^+e^-\to q\bar q$ 
component by $\pm 4.0\%$ according to the known cross section~\cite{Ablikim:2007zz}, and varying the size of the $D^+\to K^0_L\pi^+\pi^+\pi^-$ background by 30\%, which is the  largest data-simulation difference of $K^0_L$ interaction in the EMC~\cite{BESIII:2015jmz}. 
The uncertainties due to the $U_{\rm miss}$ resolution after the kinematic fit are studied with a control sample of $D^+\to K_S^0\pi^0 e^+\nu_e$.
The effects of the uncertainties on the BFs of the $\eta^{\prime}$ and $\eta$ decays~\cite{PDG2024} are also included.  
The uncertainty due to the limited simulation-sample statistics, dominated by the DT efficiency, is also considered as a systematic uncertainty.

The total systematic uncertainties are 4.3\% and 3.8\% for $D^+\to\eta^\prime \mu^+\nu_\mu$ and $D^+\to\eta^\prime e^+\nu_e$, respectively.

To study the $D^+\to\eta^{\prime} \ell^+\nu_\ell$ decay dynamics, the individual candidate events are grouped into four $q^2$ intervals. A least-$\chi^2$ fit is performed to the measured, $\Delta\Gamma_{\rm msr}$, and the theoretically expected, $\Delta\Gamma_{\rm th}$, differential decay rates among $q^2$ intervals with covariance matrix $C$~\cite{BESIII:2015tql}. The  $\chi^{2}$ is given by $\chi^{2} = \sum_{i,j=1}\left(\Delta\Gamma^{i}_{\mathrm{msr}}-\Delta\Gamma^{i}_{\mathrm{th}}\right) (C^{-1})_{ij}\left(\Delta\Gamma^{j}_{\mathrm{msr}}-\Delta\Gamma^{j}_{\mathrm{th}}\right)$. The indices $i$ and $j$ represent the different generated $q^2$ intervals and $C^{-1}$ is  the inverse of the covariance matrix $C$.

The $\Delta\Gamma^i_{\rm msr}$ are determined by
$\Delta \Gamma^i_{\rm msr}= \frac{N_{\rm prd}^i}{\tau_{D^+}
  \,  N^{\rm tot}_{\rm ST}}$, where
$\tau_{D^+}$ is the $D^+$ meson lifetime~\cite{PDG2024,Aaij:2017vqj}
and $N^i_{\rm prd}= \sum^{4}_{k}(\epsilon^{-1})_{ik} N_{\rm DT}^{k}$ is the corresponding produced signal yield.
Here, the observed signal yield, $N^k_{\rm DT}$, is obtained from a fit to the corresponding $U_{\rm miss}$ distribution in the $k^{\rm th}$ reconstructed $q^2$ interval. The signal efficiency matrix, $\epsilon$, includes the migration between the generated and reconstructed $q^2$ intervals, and $\epsilon^{-1}$ is the inverse matrix.
Details are shown in Table~3 of Ref.~\cite{Supplement}.

The partial rate, $\Delta\Gamma^i_{\rm th}$, 
relates to the hadronic FF via~\cite{Faustov:2019mqr}

\begin{widetext}
\begin{equation}
\begin{array}{l}
        \displaystyle\Delta\Gamma^i_{\rm th}=
  \int_{q^2_{{\rm min}, i}}^{q^2_{{\rm max}, i}}
  \frac{G_{F}^{2}|V_{cd}|^{2}}{24\pi^{3}}
  \frac{(q^{2}-m^{2}_{\ell})^2|p_{\eta^{\prime}}|}{q^{4}m^{2}_{D^+}}
  \left [ \left( 1+\frac{m^{2}_{\ell}}{2q^{2}} \right)
     m^{2}_{D^+} |p_{\eta^{\prime}}|^2 |f^{\eta^{\prime}}_{+}(q^{2})|^{2}    
  + \frac{3m^{2}_{\ell}}{8q^{2}}
  \left( m^{2}_{D^+}-m^{2}_{\eta^{\prime}} \right)^{2}
  |f^{\eta^{\prime}}_{0}(q^{2})|^{2}\right ]dq^2,
\end{array}
\end{equation}
\end{widetext}
where  $G_F$ is the Fermi coupling constant~\cite{PDG2024}, $p_{\eta^{\prime}}$ is the momentum of $\eta^{\prime}$ in the $D^+$ rest frame, $m_{\ell(\eta^{\prime})}$ is the lepton or $\eta^{\prime}$ mass~\cite{PDG2024}. The hadronic FF, $f_+^{\eta^{\prime}}(q^2)$, is parameterized with the two-parameter series expansion~\cite{Becher:2005bg}, in which the  product of $f_+^{\eta^\prime}(0)|V_{cd}|$ and the shape parameter $r_1$ are to be determined.  
Similar formulas are applied for $f_0^{\eta^{\prime}}(q^2)$ but with a one-parameter series expansion, due to the small contribution from this term, and with the pole mass taken from the nearest scalar charm meson, the $D^{*}_0(2300)$.  

\begin{figure}
\centering
  \includegraphics[width=0.475\textwidth]{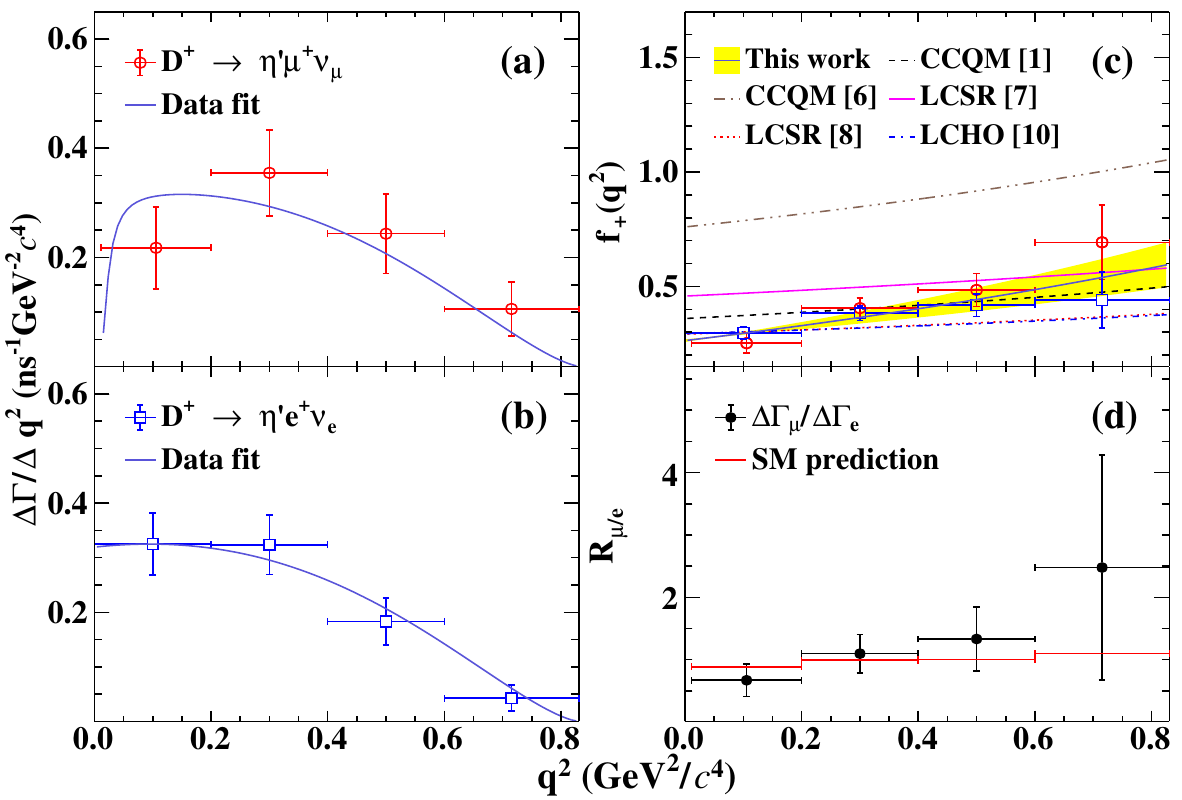}
\caption{\small (a, b) Fits to $\Delta\Gamma^i_{\rm msr}$, (c) projections to $f_+^{\eta^{\prime}}(q^2)$, and (d) the measured $\mathcal R_{\mu/e}$ in each $q^2$ interval. 
The red circles, blue square, and the black points with error bars are data; the error bars combine both statistical and systematic
uncertainties. 
The yellow bands are
the $\pm1\sigma$ intervals of the fitted parameters.}
  \label{fig:combine_FF}
\end{figure}

We perform a simultaneous fit to the differential decay rates of $D^+\to\eta^\prime \mu^+\nu_\mu$ and $D^+\to\eta^\prime e^+\nu_e$,
where the two modes are constrained to have the same parameters for the hadronic FF. 
Figures~\ref{fig:combine_FF}(a) and~\ref{fig:combine_FF}(b) exhibit the fit results and Fig.~\ref{fig:combine_FF}(c) shows the extracted hadronic FF. The goodness of fit is $\chi^2$/NDOF = 3.8/6, where NDOF is the number of degrees of freedom.  From the fit, we obtain the product of $f_+^{\eta^\prime}(0)|V_{cd}|=(5.92\pm0.56_{\rm stat}\pm0.13_{\rm syst})\times 10^{-2}$ and the shape parameter $r_1=-21.5\pm6.1_{\rm stat}\pm0.8_{\rm syst}$. The correlation coefficient between the fitted parameters is 0.88. 
The nominal fit parameters are taken from the fit with the combined statistical and systematic covariance matrix, and the statistical uncertainties of the fit parameters are taken from the fit with only the statistical covariance matrix. For each parameter, the systematic uncertainty is obtained by calculating the quadratic difference of uncertainties between these two fits. 

The statistical and systematic covariance matrices are constructed as
$C_{ij}^{\rm stat} = \left( \frac{1}{\tau_{D^+}
    N_{\rm ST}^{\rm tot}} \right)^2
\sum_{\alpha}(\epsilon^{-1})_{i\alpha}(\epsilon^{-1})_{j\alpha}
  [\sigma(N^{\alpha}_{\rm DT})]^2$ and
$C_{ij}^{\rm syst} = \delta(\Delta \Gamma^i_{\rm msr}) \, \delta(\Delta \Gamma^j_{\rm msr})$,
respectively,
where $\sigma(N^\alpha_{\rm DT})$ and $\delta(\Delta\Gamma^i_{\rm msr})$
are the statistical and systematic uncertainties in the $\alpha^{\rm th}$ and $i^{\rm th}$ $q^2$ intervals.
The sources of systematic uncertainties are almost the same as in the BF measurement, except that an additional systematic uncertainty from $\tau_{D^+}$ of 0.5\%~\cite{PDG2024} is included.
The $C_{ij}^{\rm syst}$ are obtained by summing the covariance matrices
for all systematic uncertainties.
The correlated and uncorrelated systematic uncertainties between  $D^+\to\eta^\prime \mu^+\nu_\mu$ and $D^+\to\eta^\prime e^+\nu_e$ are summarized in the top and bottom sections of Table 2 in Ref.~\cite{Supplement}, and 
the systematic uncertainty of $\tau_{D^+}$ is correlated.
The correlation matrix $\rho_{ij}=\frac{C_{ij}}{\sqrt{C_{ii}C_{jj}}}$ of the obtained $C_{ij}^{\rm stat}$ and $C_{ij}^{\rm syst}$ for different signal decays are shown in
Table 4 of Ref.~\cite{Supplement}.
The final $C_{ij}$ is obtained as $C_{ij}=C_{ij}^{\rm stat}+C_{ij}^{\rm syst}$.

The ratio of $\mathcal B(D^+\to\eta^\prime \mu^+\nu_\mu)$ to $\mathcal B(D^+\to\eta^\prime e^+\nu_e)$ is ${\mathcal R}_{\mu/e}=1.07\pm0.19_{\rm stat}\pm 0.03_{\rm syst}$,
which is consistent with the SM predictions of LFU~\cite{Soni:2018adu,Cheng:2017pcq}.
In addition, we examine ${\mathcal R}_{\mu/e}$
in different $q^2$ intervals after considering the correlated uncertainties, with results shown in Fig.~\ref{fig:combine_FF}(d); these are also consistent with the SM predictions.

\begin{figure}
\centering
  \includegraphics[width=0.475\textwidth]{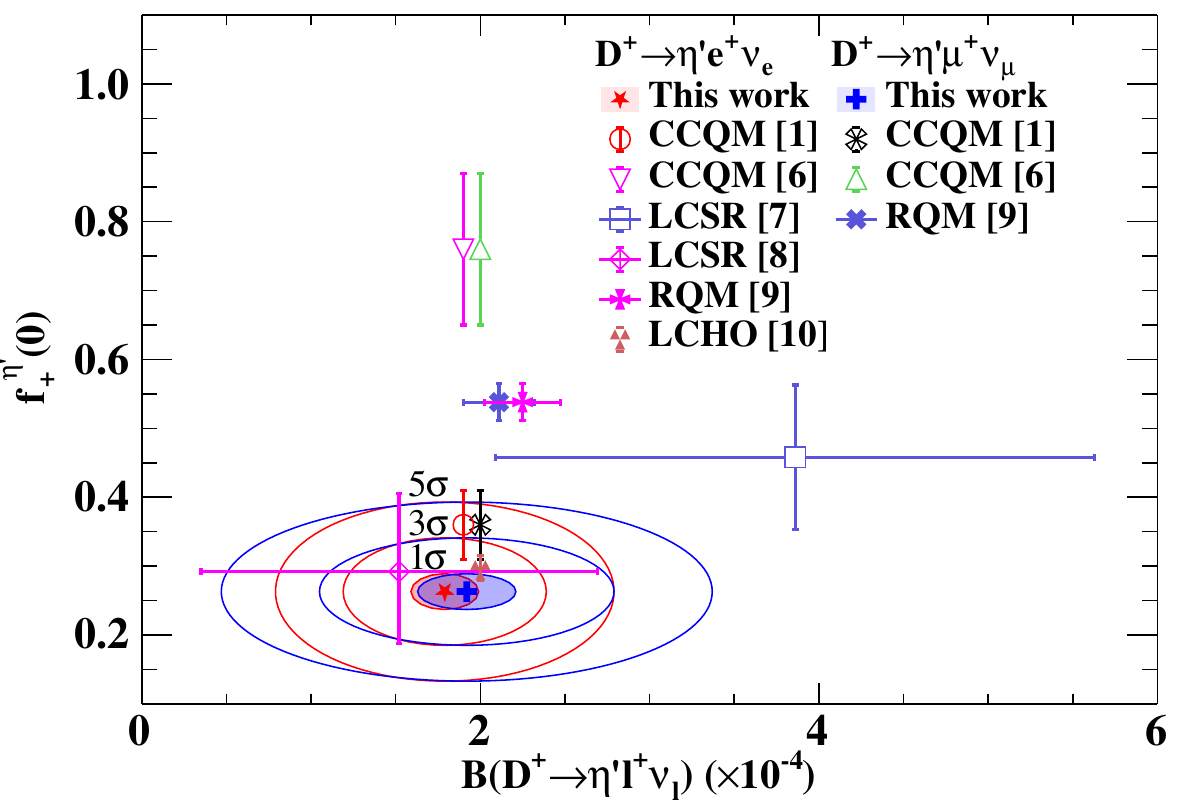}
\caption{\small Comparison of the BFs and FFs measured in this work and those given by different theoretical predictions.}
  \label{fig:com_BF_FF}
\end{figure}

To summarize, 
this Letter reports the first observation of $D^+\to \eta^\prime \mu^+\nu_\mu$ and an improved measurement of $D^+\to \eta^\prime e^+\nu_e$. 
By analyzing the $D^+\to \eta^{\prime} \ell^+\nu_\ell$ decay dynamics, 
we obtain $f_+^{\eta^{\prime}}(0)|V_{cd}|=(5.92\pm0.56_{\rm stat}\pm0.13_{\rm syst})\times 10^{-2}$ for the first time.
Taking $|V_{cd}|=0.22487\pm0.00068$ from the SM global fit~\cite{PDG2024} as input,
we determine $f^{\eta^{\prime}}_+(0)=0.263\pm0.025_{\rm stat}\pm0.006_{\rm syst}$. Figure~\ref{fig:com_BF_FF} shows the comparison of the measured and different theoretical predicted BFs and FFs. The measured results are consistent with the CCQM~\cite{Ivanov:2019nqd}, LCSR~\cite{Duplancic:2015zna,Offen:2013nma}, and LCHO~\cite{Hu:2023pdl} calculations within $2\sigma$ and disfavor the CCQM~\cite{Soni:2018adu} and RQM~\cite{Faustov:2019mqr} calculations by more than $2\sigma$.
We have also tested LFU with $D^+\to \eta^\prime \ell^+\nu_e$ in full and separate $q^2$ intervals, and no LFU violation is found.  
Combining with the BFs of $D^+ \to\eta \ell^+ \nu_\ell$~\cite{PDG2024} and
$D^+_s \to\eta^{(\prime)} \ell^+ \nu_\ell$~\cite{BESIII:2023gbn,BESIII:2023ajr}, we determine the $\eta-\eta^\prime$ mixing angle $\phi_{\rm P}=(39.8\pm0.8_{\rm stat}\pm0.3_{\rm syst})^\circ$ after considering the (un-)correlated uncertainties, which provides complementary data to constrain the gluon component in the $\eta^\prime$ meson.

The BESIII Collaboration thanks the staff of BEPCII and the IHEP computing center for their strong support. This work is supported in part by National Key R\&D Program of China under Contracts Nos. 2023YFA1606000, 2020YFA0406400, 2020YFA0406300; National Natural Science Foundation of China (NSFC) under Contracts Nos. 12305089, 11635010, 11735014, 11935015, 11935016, 11935018, 12025502, 12035009, 12035013, 12061131003, 12192260, 12192261, 12192262, 12192263, 12192264, 12192265, 12221005, 12225509, 12235017, 12361141819; the Chinese Academy of Sciences (CAS) Large-Scale Scientific Facility Program; the CAS Center for Excellence in Particle Physics (CCEPP); Joint Large-Scale Scientific Facility Funds of the NSFC and CAS under Contract Nos. U1932102, U1832207; 100 Talents Program of CAS; Jiangsu Funding Program for Excellent Postdoctoral Talent under Contract No. 2023ZB833; Project funded by China Postdoctoral Science Foundation under Contract No. 2023M732547; The Institute of Nuclear and Particle Physics (INPAC) and Shanghai Key Laboratory for Particle Physics and Cosmology; German Research Foundation DFG under Contracts Nos. 455635585, FOR5327, GRK 2149; Istituto Nazionale di Fisica Nucleare, Italy; Ministry of Development of Turkey under Contract No. DPT2006K-120470; National Research Foundation of Korea under Contract No. NRF-2022R1A2C1092335; National Science and Technology fund of Mongolia; National Science Research and Innovation Fund (NSRF) via the Program Management Unit for Human Resources \& Institutional Development, Research and Innovation of Thailand under Contract No. B16F640076; Polish National Science Centre under Contract No. 2019/35/O/ST2/02907; The Swedish Research Council; U. S. Department of Energy under Contract No. DE-FG02-05ER41374.

\clearpage
\appendix
\twocolumngrid
\setcounter{table}{0}
\setcounter{figure}{0}

\section*{Appendices}\label{Supplement}

\begin{figure*}[htp]
\centering
\includegraphics[width=0.9\textwidth]{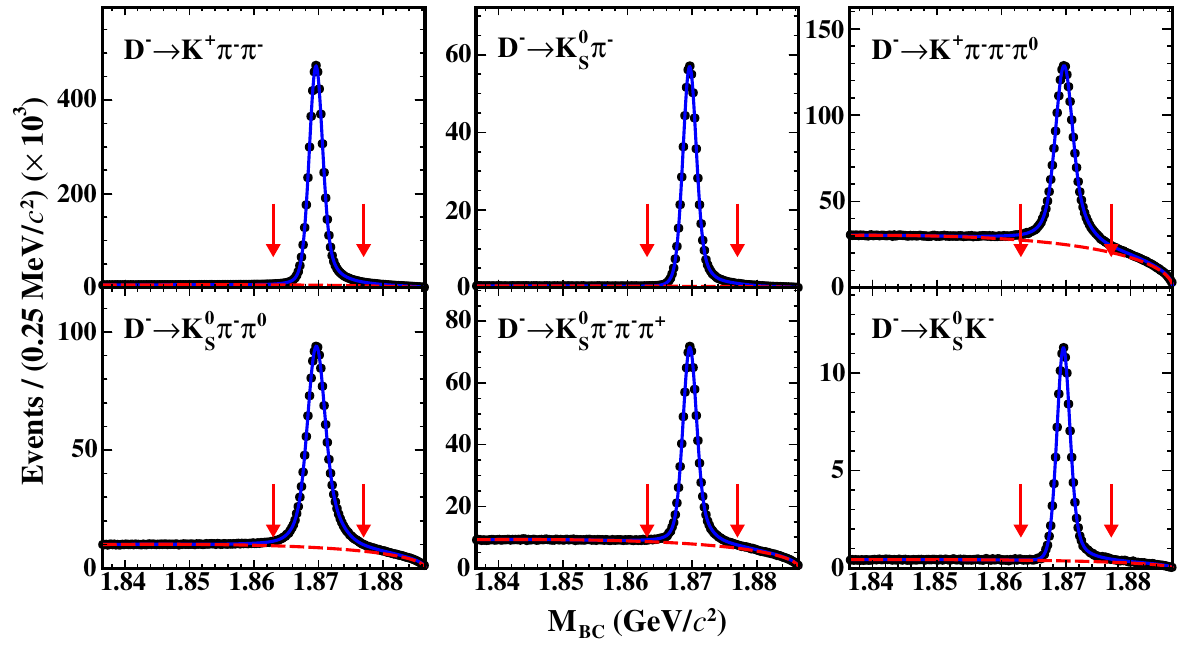}
\caption{Fits to the $M_{\rm BC}$ distributions of the ST $D^-$ candidates. In each plot, the points with error bars correspond to the data, the blue curves are the best fits, and the red dashed curves describe the fitted combinatorial background shapes. The pair of red arrows indicate the $M_{\rm BC}$ signal window.
\label{fig:STfit}}
\end{figure*}

Figure~\ref{fig:STfit} shows the fits to the $M_{\rm BC}$ distributions of the accepted ST candidates in data for different ST modes. Table~\ref{datasignum} summarizes the $\Delta E$ and $M_{\rm BC}$ requirements, the measured ST $D^-$ yields in the data, and the ST efficiencies for six tag modes.

Table~\ref{tab:sys} summarizes the sources of systematic uncertainty in the measurements of the branching fractions of $D^+\to\eta^{\prime}\ell^+\nu_\ell$. In this table, the contributions listed in the top part are treated as correlated, while those in the bottom part are uncorrelated.

 Table~\ref{tab:decayrateb} summarizes the measured partial widths in  different reconstructed $q^2$ intervals for both $D^+\to \eta^\prime \mu^+\nu_\mu$ and $D^+\to \eta^\prime e^+\nu_e$.

Table ~\ref{tab:covb} presents the statistical and systematic correlation matrices as well as the relative uncertainties for the measured partial decay
rates in different $q^2$ intervals for $D^+\to \eta^\prime \ell^+\nu_\ell$.

\begin{table*}[htp]
\centering
\caption{The $\Delta E$ and $M_{\rm BC}$ requirements, the measured ST $D^-$ yields in the data ($N_{\rm ST}^i$), and the ST efficiencies ($\epsilon_{\rm ST}^i$) for six tag modes. The uncertainties are statistical only.\label{datasignum}}
\begin{tabular}{lcccc}
  \hline
  \hline
  Tag mode            & $\Delta E$ (GeV)       &   $M_{\rm BC}$ (GeV/$c^2$)               &  $N_{\rm ST}^i$ ($\times 10^3$)&$\epsilon_{\rm ST}^i$ (\%)\\
  \hline
  $D^-\to K^+\pi^-\pi^-$&(-0.025,\,0.024)&(1.863,\,1.877)&$5552.8\pm2.5$&51.10$\pm$0.01\\
$D^-\to K_S^0\pi^-$& (-0.025,\,0.026)&(1.863,\,1.877) &$656.5\pm0.8$&51.42$\pm$0.01\\
$D^-\to K^+\pi^-\pi^-\pi^0$ &(-0.057,\,0.046)&(1.863,\,1.877) & $1723.7\pm1.8$& 24.40$\pm$0.01\\
$D^-\to K_S^0\pi^-\pi^0$ &(-0.062,\,0.049)&(1.863,\,1.877) & $1442.4\pm1.5$&26.45$\pm$0.01\\
$D^-\to K_S^0\pi^+\pi^-\pi^-$ &(-0.028,\,0.027)&(1.863,\,1.877) & $790.2\pm1.1$& 29.59$\pm$0.01\\
$D^-\to K_S^0K^-$ &(-0.024,\,0.025)&(1.863,\,1.877) & $124.3\pm0.4$ & 47.82$\pm$0.02\\
      \hline
  \hline
\end{tabular}
\end{table*}

\begin{table*}[htp]
\centering
\caption{Relative systematic uncertainties (in \%) in the measurement of the BFs
of $D^+\to\eta^\prime\ell^+\nu_\ell$. The top and bottom parts are correlated and uncorrelated, respectively.  
Items with a $-$ are negligible.  \label{tab:sys}
}
\begin{tabular}{lcc}
  \hline
  \hline
 Source  &$\eta^\prime \mu^+\nu_\mu$&$\eta^\prime e^+\nu_e$  \\
  \hline
  $N^{\rm tot}_{\rm ST}$                &0.3              &0.3      \\
   $\pi^\pm$ tracking                    &1.0 &1.0\\
  $\pi^\pm$ PID                            &1.0 &1.0\\
  Photon and $\eta$  reconstruction   &1.0               &1.0\\
  $\eta^\prime$ reconstruction      &1.0                &1.0\\
    $\cos\theta_{\pi, \rho}$ requirement      & $-$               &$-$         \\
  $E^{\rm max}_{\rm extra \gamma}\&N_{\rm extra}^{\rm char}$\&$N_{\rm extra}^{\pi^0}$ &2.1               &2.0    \\
  Signal model                               &1.2                &1.2      \\
  $M_{\rm rec}^2$ requirement      & 1.8               &1.7         \\
 Quoted BF                                 &1.0               &1.0\\
      \hline

         $\ell^\pm$ tracking                   &0.5                &0.5        \\
  $\ell^\pm$ PID                          & 1.0                &1.0        \\

                      $M_{\rm \eta^{\prime}\ell^+}$ requirement  &$-$&   $-$       \\
                                                      $\chi^2$     requirement                    &1.0      &    $-$                \\
            $\cos\theta_{b}$ requirement &1.0  &$-$         \\
 $U_{\rm miss}$ fit              &1.1        &0.4                         \\
  $U_{\rm miss}$ resolution                     &$-$                &$-$         \\
  MC  statistics                           &0.2                &0.2         \\
  \hline
  Total                                &4.3            &3.8           \\

  \hline
  \hline
\end{tabular}

\end{table*}

\begin{table*}[htp]\centering
\caption{
The partial decay rates of $D^+\to\eta^\prime \ell^+\nu_\ell$ in different $q^{2}$ intervals. The uncertainties are statistical only. }
\label{tab:decayrateb}
\begin{tabular}{cccccc}\hline\hline
\multicolumn{2}{c}{$i$}&1&2&3&4\\
\multicolumn{2}{c}{$q^2$ $(\mathrm{GeV}^{2}/c^{4})$}&($m_\ell^2,\,0.20$)&($0.20,\,0.40$)&($0.40,\,0.60$)&($0.60,\,0.83$)\\
\hline
\multirow{3}{*}{$\mu$}&$N_{\mathrm{DT}}^i$&20.5$\pm$6.8&35.4$\pm$7.6&22.7$\pm$6.4&10.0$\pm$4.4\\
&$N_{\mathrm{prd}}^i$&438$\pm$151&757$\pm$168&519$\pm$156&260$\pm$122\\
&$\Delta\Gamma^i_{\rm msr}$ $(\mathrm{ns^{-1}})$&0.041$\pm$0.014&0.071$\pm$0.016&0.049$\pm$0.015&0.024$\pm$0.011\\
\hline
\multirow{3}{*}{$e$}&$N_{\mathrm{DT}}^i$&54.0$\pm$8.9&55.7$\pm$8.9&33.9$\pm$7.5&9.7$\pm$4.9\\
&$N_{\mathrm{prd}}^i$&693$\pm$120&690$\pm$116&390$\pm$92&105$\pm$58\\
&$\Delta\Gamma^i_{\rm msr}$ $(\mathrm{ns^{-1}})$&0.065$\pm$0.011&0.065$\pm$0.011&0.037$\pm$0.009&0.010$\pm$0.005\\

\hline\hline
\end{tabular}
\end{table*}

\begin{table*}[htp]\centering
\caption{Statistical and systematic correlation matrices as well as the relative uncertainties of the measured partial decay rate in each
 for the measured partial decay rates of  $D^+\to\eta^\prime \ell^+\nu_\ell$ in different
$q^2$ intervals. }
\label{tab:covb}
\begin{tabular}{ccccccccc}\hline\hline
\multicolumn{9}{c}{Statistical correlation matrix}\\
\multirow{2}{*}{$\rho_{ij}^{\rm stat}$}&\multicolumn{4}{c}{$D^+\to\eta^\prime \mu^+\nu_\mu$}&\multicolumn{4}{c}{$D^+\to\eta^\prime e^+\nu_e$}\\
&1&2&3&4&1&2&3&4\\
\hline
1	&1.000	&$-$0.049	&$-$0.002	&$-$0.004	&0.000	&0.000	&0.000	&0.000	\\
2	&&1.000	&$-$0.055	&$-$0.002	&0.000	&0.000	&0.000	&0.000	\\
3	&&&1.000	&$-$0.053	&0.000	&0.000	&0.000	&0.000	\\
4	&&&&1.000	&0.000	&0.000	&0.000	&0.000	\\
1	&&&&&1.000	&$-$0.071	&0.003	&$-$0.001	\\
2	&&&&&&1.000	&$-$0.073	&0.002	\\
3	&&&&&&&1.000	&$-$0.072	\\
4	&&&&&&&&1.000	\\
Stat (\%)&34.4&22.2&30.0&47.0&17.3&16.9&23.5&55.5\\
\hline 
\multicolumn{9}{c}{Systematic correlation matrix}\\
\multirow{2}{*}{$\rho_{ij}^{\rm syst}$}&\multicolumn{4}{c}{$D^+\to\eta^\prime\mu^+\nu_\mu$}&\multicolumn{4}{c}{$D^+\to\eta^\prime e^+\nu_e$}\\
&1&2&3&4&1&2&3&4\\
\hline
1	&1.000	&0.837	&0.797	&0.779	&0.770	&0.747	&0.685	&0.506	\\
2	&&1.000	&0.969	&0.913	&0.883	&0.892	&0.863	&0.663	\\
3	&&&1.000	&0.932	&0.839	&0.879	&0.891	&0.704	\\
4	&&&&1.000	&0.820	&0.841	&0.830	&0.645	\\
1	&&&&&1.000	&0.979	&0.913	&0.669	\\
2	&&&&&&1.000	&0.964	&0.719	\\
3	&&&&&&&1.000	&0.815	\\
4	&&&&&&&&1.000	\\
Syst (\%)&4.9&4.1&4.1&4.3&3.9&3.8&3.9&5.0\\
\hline\hline
\end{tabular}
\end{table*}


\begin{thebibliography}{99}
\bibitem{charge}
Inclusion of charge-conjugate modes is implied throughout this Letter.
\end{thebibliography}
\end{document}